\documentclass[aps,prd,preprint,tightenlines,groupedaddress,showpacs,nofootinbib]{revtex4}
\usepackage{epsf,epsfig,graphics,graphicx,amsmath}
\begin{document}

\title{Superhorizon curvature perturbation in ultra-slow-roll inflation}

\author{Shu-Lin Cheng$^1$}
\author{Wolung Lee$^1$}
\author{Kin-Wang Ng$^{2,3}$}

\affiliation{
$^1$Department of Physics, National Taiwan Normal University,
Taipei 11677, Taiwan\\
$^2$Institute of Physics, Academia Sinica, Taipei 11529, Taiwan\\
$^3$Institute of Astronomy and Astrophysics, Academia Sinica, Taipei 11529, Taiwan}

\date{\today}

\begin{abstract}
We study the growth of superhorizon modes in the curvature perturbation during an ultra-slow-roll or a large-$\eta$ phase in single-field inflation. In a simple toy model, we derive the two-point correlation function of the curvature perturbation and show that the requirement for causality restricts the growth rate and hence puts a lower limit on the value of $\eta$. The toy model is then realized by considering an inflation potential with an inflection point. Our study is useful to assessing the growth of the curvature perturbation that seeds the formation of primordial black holes.
\end{abstract}

\maketitle

\section{Introduction}

The inflation scenario is generally accepted for explaining the observed spatial flatness and homogeneity of the Universe.
A simple model of the scenario such as the slow-roll inflation driven by a flat inflaton potential predicts quasi de Sitter
vacuum fluctuations. These fluctuations could give rise to Gaussian and nearly scale-free metric perturbation containing both density fluctuations and gravitational waves ~\cite{olive}. 

Recent cosmological observations have measured the density perturbation in an unprecedented accuracy, thus supporting the slow-roll inflation model~\cite{planckinflation}. Future observations will target at unveiling the detailed features of the metric perturbation such as non-gaussianities, the running spectral index, and the B-modes~\cite{cmb-s4}. These large-scale measurements have precisely determined the slow-roll parameters of the inflaton potential in the beginning of inflation or about 60 e-folds before the end of inflation; however, they can hardly tell us anything about inflation in a later stage or at smaller scales. 

Recently, stimulated by the detection of gravitational waves from binary-black-hole mergers made by Advanced LIGO/VIRGO~\cite{LIGO}, various inflation models have been re-visited to explore the production of large curvature perturbation in a later stage of the inflationary epoch that seeds the formation of primordial black holes~\cite{pbhs,inflection}. In particular, a specific model of the single-field inflation can produce a large curvature perturbation as long as the potential has an inflection point at which the slow-roll condition is violated~\cite{inflection}. In this paper, we will scrutiny this model, concentrating on the time evolution of the superhorizon modes of the curvature perturbation.

\section{Single-field inflation model}

In a flat universe with a scale factor $a$, the equation of motion of the inflaton in a single-field inflation model with a given potential $V(\phi)$ is given by
\begin{equation}
\ddot{\phi} + 3H\dot{\phi} + \frac{dV}{d\phi} = 0,
\label{phieom}
\end{equation}
where the dot denotes the derivative with respect to the time $t$ and the Hubble parameter $H=\dot{a}/a$ is governed by the Friedmann equation,
\begin{equation}
H = \sqrt{\frac{1}{3}\left( \frac{1}{2} \dot{\phi}^2 + V \right)},
\label{heom}
\end{equation}
in which we have set the reduced Planck mass $M_p=1$. Let us introduce the Hubble flow parameters
\footnote{Originally $\epsilon$ and $\eta$ are termed as the Hubble slow-roll parameters. On cosmological scales, observations constrain them to satisfy the so-called slow-roll condition: $|\epsilon|\ll 1$ and $|\eta|\ll 1$. Here we discuss a situation with arbitrary values of $\epsilon$ and $\eta$, so for our purpose we rename them as the Hubble flow parameters.},
\begin{equation}
\epsilon = - \frac{\dot{H}}{H^2},\quad \eta = \frac{\dot\epsilon}{\epsilon H}.
\end{equation}
From Eqs.~(\ref{phieom}) and (\ref{heom}), they become
\begin{equation}
\epsilon = \frac{\dot{\phi}^2}{2 H^2},\quad \eta= -6 - \frac{2}{\dot\phi H} \frac{dV}{d\phi} + \frac{\dot\phi^2}{H^2}.
\end{equation}

In the course of inflation, quantum fluctuations of the inflaton field give rise to the curvature perturbation. In terms of the Mukhanov-Sasaki variable $v$~\cite{msvar}, the Fourier mode $v_k$ satisfies
\begin{equation}
v_k'' +\left(k^2-\frac{z''}{z}\right) v_k = 0,
\label{MS1}
\end{equation}
where the prime denotes the conformal time derivative, $d\tau=dt/a$, and $z=a\sqrt{\epsilon}$.
Here the initial condition for $v_k$ is chosen to be the standard Bunch-Davies vacuum state:
\begin{equation}
v_k\to \frac{1}{\sqrt{2k}} e^{-ik\tau},\quad\text{as}\;\frac{k}{aH} \gg 1.
\label{initialcon}
\end{equation}
We can re-write Eq.~(\ref{MS1}) in terms of $\zeta_k=v_k/(a\sqrt{2\epsilon})$ as
\begin{equation}
\zeta_k'' +\left(2 + \eta \right)aH \zeta_k' + k^2  \zeta_k = 0,
\label{MS2}
\end{equation}
and the power spectrum of the curvature perturbation is given by
\begin{equation}
\Delta^2_{\zeta_k}=\frac{k^3}{2\pi^2} \left| \zeta_k \right|^2.
\label{deltazetak}
\end{equation}

For a slow-roll inflation with $|\epsilon|\ll 1$ and $|\eta|\ll 1$,  $H$ and $\epsilon$ are approximately constant and we have $a=e^{Ht}=-1/(H\tau)$. Therefore, Eq.~(\ref{MS2}) has a simple analytic solution,
\begin{equation}
\zeta_k= \frac{\pi^{1/2}H}{2k^{3/2}}\frac{1}{\sqrt{2\epsilon}}(-k\tau)^{3/2} H^{(1)}_{3/2}(-k\tau),
\end{equation}
where only the Hankel function of the first kind appears because $\sqrt{2\epsilon}\zeta_k \to e^{-ik\tau}/(a\sqrt{2k})$ as $-k\tau\gg 1$. Further, the power spectrum of the curvature perturbation can be found by taking the late time limit ($-k\tau\ll 1$) as
\begin{equation}
\Delta^2_{\zeta_k}=\frac{H^2}{8\pi^2 \epsilon}.
\label{deltaappro}
\end{equation}

\section{A toy model with a large $\eta$}
\label{toy}

Let us consider a toy inflation model that has $|\epsilon|\ll 1$ and the temporal change of $\eta$ given by
\begin{equation}
    \eta=
    \begin{cases}
      0, & \quad\ \tau_0<\tau<\tau_1 \\
     {\bar\eta} , & \quad\ \tau_1<\tau<\tau_2 \\
      0, & \quad\ \tau_2<\tau<\tau_e 
    \end{cases}
\end{equation}
where $\tau_0$ and $\tau_e$ denote the beginning and the end of inflation, respectively. Then, $H$ can be treated as a constant, $a=e^{Ht}=-1/(H\tau)$, and Eq.~(\ref{MS2}) has the solution,
\begin{equation}
   A^{-1} \zeta_k=
    \begin{cases}
      (-k\tau)^{3/2} H^{(1)}_{3/2}(-k\tau), & \quad\ \tau_0<\tau<\tau_1 \\
      (-k\tau)^\nu \left[c_1H^{(1)}_\nu(-k\tau)+c_2H^{(2)}_\nu(-k\tau)\right], & \quad\ \tau_1<\tau<\tau_2 \\
      (-k\tau)^{3/2} \left[d_1H^{(1)}_{3/2}(-k\tau)+d_2H^{(2)}_{3/2}(-k\tau)\right], & \quad\ \tau_2<\tau<\tau_e 
    \end{cases}
\label{zetasoln}
\end{equation}
where 
\begin{equation}
A=\pi^{1/2}H/(2k^{3/2}\sqrt{2\epsilon_0}),\quad \nu=(3+{\bar\eta})/2. 
\end{equation}
The power spectrum of the curvature perturbation at the end of inflation for modes satisfying $-k\tau_e\ll 1$ is simply given by
\begin{equation}
\Delta^2_{\zeta_k}=\frac{H^2}{8\pi^2 \epsilon_0} |d_1-d_2|^2.
\label{deltad1d2}
\end{equation}

Given a $k$-mode, we can match the above piecewise functions at $\tau=\tau_1$ and $\tau=\tau_2$ to determine $c_1$, $c_2$, $d_1$, and $d_2$. First of all, it is interesting to look at a special case in which ${\bar\eta}=-6$ ($\nu=-3/2$). Using the relationships,
\begin{equation}
H^{(1)}_{-\alpha}(x)=e^{i\alpha\pi}H^{(1)}_{\alpha}(x)\quad\text{and}\quad
H^{(2)}_{-\alpha}(x)=e^{-i\alpha\pi}H^{(2)}_{\alpha}(x),
\end{equation}
we find that
\begin{equation}
c_1=i(-k\tau_1)^3,\quad c_2=0;\quad d_1=\left(\frac{\tau_2}{\tau_1}\right)^{-3},\quad d_2=0.
\label{case0}
\end{equation}
For an arbitrary ${\bar\eta}$, since we are concerned with the mode growth on superhorizon scales, we discuss three representative cases as follows. 

Firstly, we consider the modes that have left the horizon before $\tau=\tau_1$ (i.e. $-k\tau_1\ll 1$).
Using the asymptotic forms at small $x$ when $\text{Re}(\alpha)>0$,
\begin{equation}
H^{(1)}_\alpha(x)\sim -H^{(2)}_{\alpha}(x)\sim - \left({i\over\pi}\right)\Gamma(\alpha) \left(x\over 2\right)^{-\alpha},
\end{equation}
we obtain that
\begin{eqnarray}
c_1-c_2\simeq\frac{\sqrt{2\pi}}{\Gamma(\nu)}2^{-\nu}, 
\quad d_1-d_2\simeq 1&&\quad\text{for}\quad\nu>0;\\
c_1 e^{-i\nu\pi} - c_2 e^{i\nu\pi} \simeq \frac{\sqrt{2\pi}}{\Gamma(-\nu)} 2^\nu (-k\tau_1)^{-2\nu},\quad
d_1-d_2\simeq \left(\frac{\tau_2}{\tau_1}\right)^{2\nu}&&\quad\text{for}\quad\nu<0. 
\label{amplify}
\end{eqnarray}
The former case (i.e. ${\bar\eta}>-3$) means that the modes with $-k\tau_1\ll 1$ are frozen once they cross out the horizon. In the latter case, they still grow even after exiting the horizon.

Next we consider a $k$-mode such that $-k\tau_1\gg1$ and $-k\tau_2\ll 1$. Using the asymptotic forms at large $x$,
\begin{eqnarray}
H^{(1)}_{\alpha}(x)&\sim&\sqrt{\frac{2}{\pi x}}e^{i\left(x-{1\over2}\alpha\pi-{1\over4}\pi\right)},\\
H^{(2)}_{\alpha}(x)&\sim&\sqrt{\frac{2}{\pi x}}e^{-i\left(x-{1\over2}\alpha\pi-{1\over4}\pi\right)},
\end{eqnarray}
we obtain that
\begin{eqnarray}
c_1\simeq -(-k\tau_1)^{{3\over2}-\nu} e^{i\left({1\over2}\nu\pi+{1\over4}\pi\right)},\quad c_2=0;&& \label{case2}\\
d_1-d_2\simeq -\frac{2^\nu}{\sqrt{2\pi}} \Gamma(\nu) (-k\tau_1)^{{3\over2}-\nu} e^{i\left({1\over2}\nu\pi+{1\over4}\pi\right)}\quad\text{for}\quad\nu>0;&&\\
d_1-d_2\simeq  -\frac{2^{-\nu}}{\sqrt{2\pi}} \Gamma(-\nu) (-k\tau_1)^{{3\over2}+\nu} \left(\frac{\tau_2}{\tau_1}\right)^{2\nu}
e^{i\left(-{1\over2}\nu\pi+{1\over4}\pi\right)}
\quad\text{for}\quad\nu<0.&&
\end{eqnarray}

Lastly, for the $k$-mode with $-k\tau_1\gg1$ and $-k\tau_2\gg1$, we find that
\begin{eqnarray}
c_1\simeq -(-k\tau_1)^{{3\over2}-\nu} e^{i\left({1\over2}\nu\pi+{1\over4}\pi\right)},\quad c_2=0;&&\label{case3}\\
d_1\simeq \left(\frac{\tau_2}{\tau_1}\right)^{\nu-{3\over2}},\quad d_2=0.&&
\end{eqnarray}

When ${\bar\eta}=0$, in all cases we have $|d_1-d_2|^2\simeq 1$ and thus the power spectrum in Eq.~(\ref{deltad1d2}) becomes the standard slow-roll solution. When ${\bar\eta}>0$, the power spectrum is suppressed. When ${\bar\eta}<0$, due to the mode growth on superhorizon scales, the power spectrum is amplified, for example in Eq.~(\ref{amplify}), by a factor, 
\begin{equation}
\left(\frac{\tau_2}{\tau_1}\right)^{4\nu} = e^{-4\nu\Delta N},
\label{amplifyN}
\end{equation}
where the e-folding is defined by $N=\int_0^t H(t')dt'$ and $\Delta N=N_2-N_1$ is the change of e-folds from $\tau_1$ to $\tau_2$.

\section{Two-point correlation function}

One may ask whether the superhorizon growth violates the causality. To answer this question, we examine the correlation length by computing the equal-time two-point correlation function. In order to save the causality, the correlation length should always be smaller than the causal horizon.

The two-point correlation function is defined by
\begin{equation}
\langle\zeta({\vec x},\tau) \zeta({\vec x}',\tau') \rangle 
= \int \frac{d^3{\vec k}}{(2\pi)^3} \zeta_k(\tau) \zeta_k^*(\tau') e^{i{\vec k}\cdot({\vec x}-{\vec x}')},
\label{2ptfct}
\end{equation}
where $\zeta_k$ is the solution in Eq.~(\ref{zetasoln}).  It suffices to consider the function in the period $\tau_1<\tau<\tau_2$ when ${\bar\eta}\neq 0$. The results in the previous section suggest that we can take $c_1$ of the form $c_1= Ck^\beta$ and set $c_2=0$. Then, Eq.~(\ref{2ptfct}) becomes
\begin{eqnarray}
&&\langle\zeta({\vec x},\tau) \zeta({\vec x}',\tau') \rangle \nonumber \\
&=&  \frac{\pi H^2}{8\epsilon_0}  |C|^2 (\tau\tau')^{\nu}
\int \frac{d^3{\vec k}}{(2\pi)^3} k^{2(\beta+\nu)-3} H^{(1)}_\nu(-k\tau) H^{(1)*}_\nu(-k\tau')
e^{i{\vec k}\cdot({\vec x}-{\vec x}')} \nonumber \\
&=&  \frac{H^2}{4\pi^3\epsilon_0}  |C|^2 \frac{(\tau\tau')^{\nu}}{\left|{\vec x}-{\vec x}'\right|}
\int_0^\infty dk\, k^{2(\beta+\nu)-2} \sin\left(k\left|{\vec x}-{\vec x}'\right|\right)
K_\nu(ik\tau) K_\nu(-ik\tau'). \nonumber
\end{eqnarray}
In most of the cases, as found in Eqs.~(\ref{case0}), (\ref{case2}), and (\ref{case3}), 
we have $|C|^2=|\tau_1|^{3-2\nu}$ and $\beta=3/2-\nu$. Hence, we obtain that
\begin{eqnarray}
&&\langle\zeta({\vec x},\tau) \zeta({\vec x}',\tau') \rangle \nonumber \\
&=&  \frac{H^2}{32\pi^2\epsilon_0} \left(\frac{\tau\tau'}{\tau_1^2}\right)^{\nu-{3\over2}}
\Gamma\left({3\over2}+\nu\right) \Gamma\left({3\over2}-\nu\right) 
{_2}F_1\left({3\over2}+\nu,{3\over2}-\nu;2;1-{1\over w}\right),\label{hyper}
\end{eqnarray}
where $|\nu|<3/2$ and
\begin{equation}
w=\frac{4\tau\tau'}{-(\tau-\tau')^2+\left|{\vec x}-{\vec x}'\right|^2}\,.
\nonumber
\end{equation}
For any two spatial points separated by a distance much greater than the horizon, i.e. $\left|{\vec x}-{\vec x}'\right|\gg |\tau|$, we can set $\tau=\tau'$ and take the limit $w\rightarrow 0^+$ in Eq.~(\ref{hyper}) to obtain the equal-time correlation,
\begin{equation}
\langle\zeta({\vec x},\tau) \zeta({\vec x}',\tau) \rangle =  
\frac{H^2}{8\pi^{5/2}\epsilon_0} \left(\frac{\tau^2}{\tau_1^2}\right)^{\nu-{3\over2}}
\left(\frac{\tau^2}{\left|{\vec x}-{\vec x}'\right|^2}\right)^{{3\over2}-|\nu|} 
 \Gamma\left(|\nu|\right) \Gamma\left({3\over2}-|\nu|\right),
\end{equation}
which diminishes with the two-point separation when $|\nu|<3/2$ ($0>{\bar\eta}>-6$). 
This means that although the modes are growing on superhorizon scales, 
the growth is causal in the sense that the correlation length is safely smaller than the horizon. 
It is worth noting that by means of the toy model, we have shown that the causality condition requires that 
${\bar\eta} \gtrsim -6$. In the next section, we will use a realistic inflation model to further examine this lower limit.

%Fig.1
\begin{figure}[htp]
\centering
\includegraphics[width=0.8\textwidth]{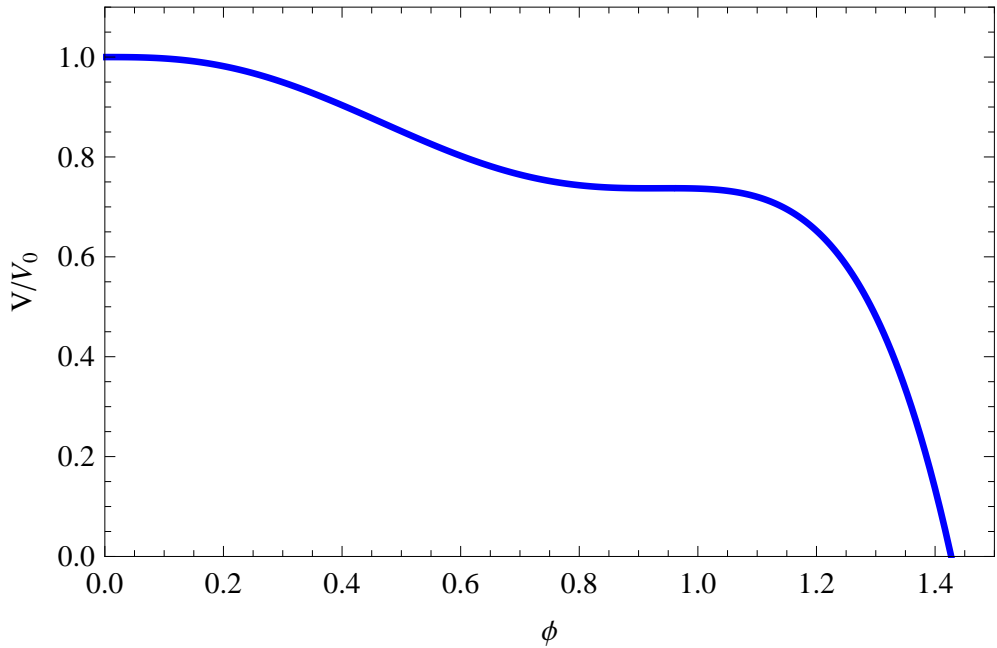}
\caption{The inflaton potential $V(\phi)$ in Eq.~(\ref{infpot}), normalized by $V_0$. All dynamical variables in this figure and in the following figures are rescaled by the reduced Planck mass, $M_p=2.435\times 10^{18}$ GeV. }
\label{fig1}
\end{figure}

%Fig.2
\begin{figure}[htp]
\centering
\includegraphics[width=0.8\textwidth]{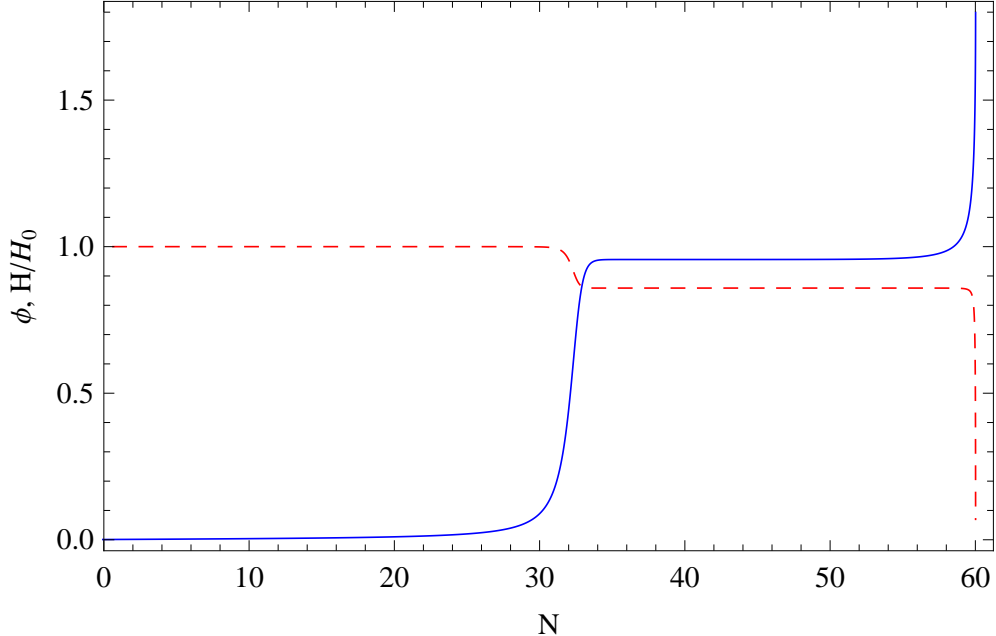}
\caption{Evolution of $H$ (dashed line) and $\phi$ (solid line) against e-folds $N$, with 
$H_0=7.65\times10^{-8}$, $\phi_0=8.48\times 10^{-4}$, and ${\dot\phi}_0=1.90\times 10^{-11}$. 
Note that the inflation ends at $N\sim 60$.}
\label{fig2}
\end{figure}

%Fig.3
\begin{figure}[htp]
\centering
\includegraphics[width=0.8\textwidth]{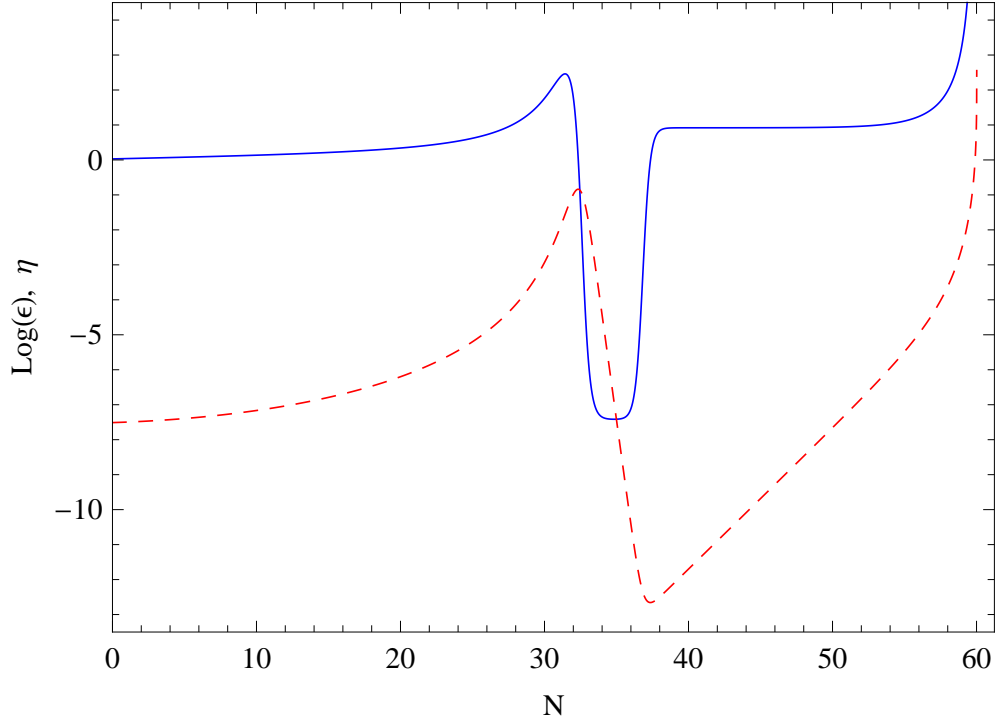}
\caption{Evolution of $\epsilon = - \dot{H}/H^2$ (dashed line) and $\eta = \dot\epsilon/(\epsilon H)$ (solid line).}
\label{fig3}
\end{figure}

%Fig.4
\begin{figure}[htp]
\centering
\includegraphics[width=0.8\textwidth]{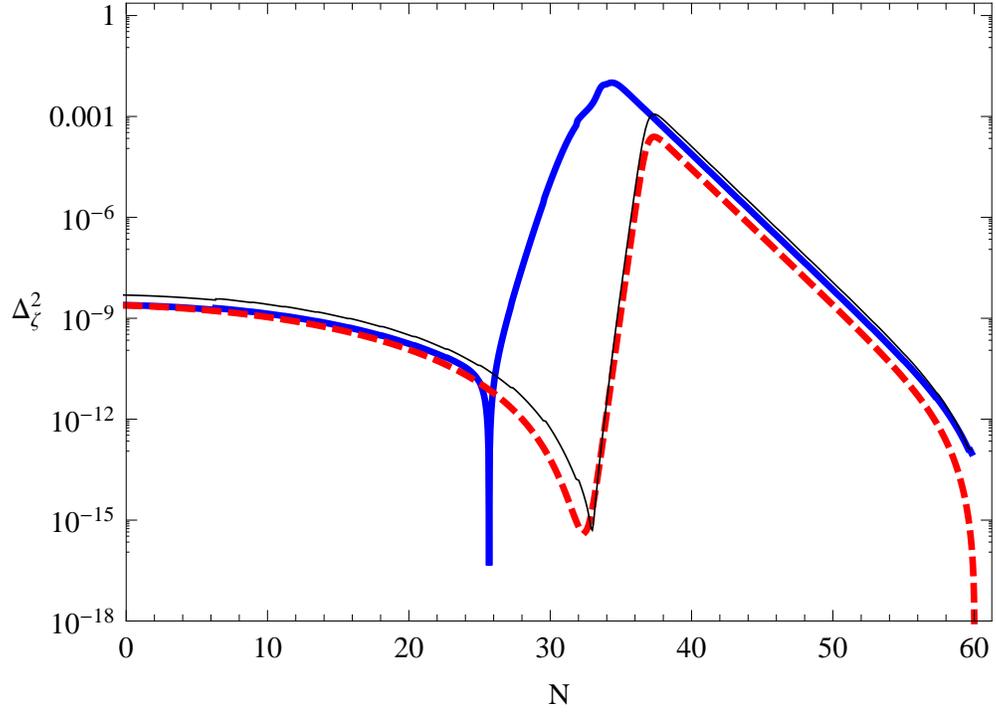}
\caption{Power spectrum of the curvature perturbation. The thick solid curve is the exact numerical result from solving Eq.~(\ref{MS2}). The dashed curve is an approximate result obtained by Eq.~(\ref{deltaappro}). The thin solid curve is the numerical result with $\zeta_k$ evaluated at the horizon-crossing time for each $k$-mode.}
\label{fig4}
\end{figure}

%Fig.5
\begin{figure}[htp]
\centering
\includegraphics[width=0.8\textwidth]{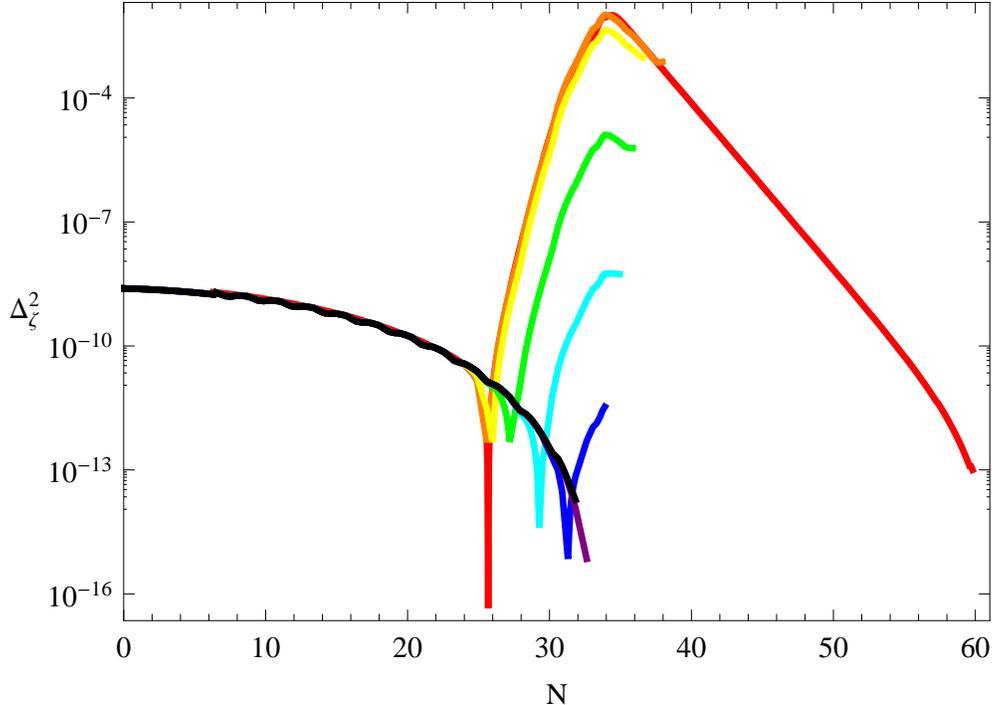}
\caption{Snapshots of the evolution of the curvature perturbation in the course of the inflation, timed by successive e-folds $N=32,33,34,35,36,37,38,60$. When drawing a curve for the snapshot at time $N$, the $k$-modes that have not yet crossed out the horizon are omitted, so the curve ends at the e-folding $N$. }
\label{fig5}
\end{figure}

\section{Inflation with an inflection point}

In order to have a large value of $\eta$, we utilize an inflation model with an inflection point. The potential is given by
\begin{equation}
V(\phi)=V_0 \left[ 1+ c_1 \frac{\phi}{c_\Lambda} + \frac{c_3}{3!} \left(\frac{\phi}{c_\Lambda}\right)^3  + \frac{c_4}{4!} \left(\frac{\phi}{c_\Lambda}\right)^4 + \frac{c_5}{5!} \left(\frac{\phi}{c_\Lambda}\right)^5 \right].
\label{infpot}
\end{equation}
We tune the above potential parameters and the initial conditions of the inflaton motion, given by
\begin{eqnarray}
&&c_\Lambda = 0.3,\quad V_0 = 1.7574 \times 10^{-14}, \nonumber \\
&&c_1 = -0.7278 \times 10^{-4},\; c_3 = -0.52,\; c_4 = 1.0,\; c_5 = -0.6407; \nonumber \\
&&\phi_0 = 8.5 \times 10^{-4},\quad {\dot\phi}_0 = 1.9 \times 10^{-11};
\end{eqnarray}
such that at near the inflection point the inflaton undergoes an ultra-slow-roll motion, while the inflaton acceleration is non-negligible. In addition, the amplitude of the power spectrum of the curvature perturbation as well as the levels of the scalar spectral index, the tensor-to-scalar ratio, and the scalar spectral index running on large cosmological scales are all consistent with the Planck measurements: 
$\Delta_{\zeta_k}^2\simeq 2.4\times 10^{-9}$, $n_s\simeq 0.97$, $r<0.06$, and $|dn_s/d\ln k|<0.013$~\cite{planckinflation,bk18}. 

Figure~\ref{fig1} is the inflaton potential with an inflection point. Figure~\ref{fig2} depicts the evolution of the Hubble parameter and the inflaton mean field. We can see that the inflaton field is almost stationary near the inflection point at $\phi\simeq 0.95$ and hence prolongs the duration of inflation by about $25$ e-folds. The evolution of the Hubble flow parameters is shown in Fig.~\ref{fig3}, where a sudden drop in $\epsilon$ at $N\sim 32$ results in a large decrease of $\eta$. The value of $\eta$ has reached about $-7$ at $N\sim 35$, which is close to the predicted lowest limit of $\eta$ by the causality argument in the previous section.

We numerically solve for $\zeta_k$ in Eq.~(\ref{MS2}) under the initial condition~(\ref{initialcon}). The power spectrum of the curvature perturbation is then given by Eq.~(\ref{deltazetak}) with $\zeta_k$ evaluated at the end of inflation. In the present consideration, we specifically compare the exact numerical result with that directly obtained by the approximate formula in Eq.~(\ref{deltaappro}). Figure~\ref{fig4} shows that the approximate formula underestimates the rapid growth of the curvature perturbation when $\eta$ has a large negative value. Also, in order to highlight the superhorizon growth, we have produced a power spectrum using Eq.~(\ref{deltazetak}) in which $\zeta_k$ takes the value when the $k$-mode crosses the horizon. This power spectrum, denoted by the thin solid curve in Fig.~\ref{fig4}, matches the one from the approximate formula. In Fig.~\ref{fig5}, we show the growth of the curvature perturbation in the course of the inflation. 

In Sec.~\ref{toy}, based on the toy model, we have shown that the growth of the superhorizon modes can boost up the power spectrum by a factor of $e^{-4\nu\Delta N}$. We can check this with the numerical results as follows. In Fig.~\ref{fig3}, the evolution profile of $\eta$ can be roughly captured by the shape of a square well with a depth of about $-7$ at $N\sim 35$ and a width of $\Delta N \simeq 3$. This gives rise to a boost factor of about $e^{24}\sim10^{12}$, which matches fairly well to the large power increase from the thin solid line to the exact power spectrum at $N\sim 34$ in Fig.~\ref{fig4}. Figure~\ref{fig5} depicts how the growth of the superhorizon modes during the short interval of a few e-folds at $N\sim 34$ renders the large boost to the power spectrum.

\section{Conclusions}

We have analyzed the evolution of the superhorizon modes of the curvature perturbation during single-field inflation in a period when the slow-roll condition is violated. The conservation of the superhorizon curvature perturbation, derived in the assumption of the slow-roll condition~\cite{conservation}, is no longer valid, as the superhorizon modes can grow rapidly in an ultra-slow-roll phase during which the speed of the inflaton is extremely low, while the acceleration is large. Nevertheless, the principle of causality implies an upper bound on the growth rate. 

We have illustrated the superhorizon growth in a full numerical calculation with the inflaton potential having an inflection point. The inflation undergoes an ultra-slow-roll phase at the inflection point such that the value of $\epsilon$ is extremely small whereas $\eta$ can reach a negative value as low as about $-7$. This value is consistent with the lower limit of $\eta\sim -6$ estimated by an analytic study in the present work. As a consequence, the curvature perturbation can grow enormously to a peak value of $\Delta^2_{\zeta_k}\sim 0.01$ at $N\sim 34$, which serves as seed for forming primordial black holes of mass about $10^{-8} M_\odot$. 

Our analytic study can be readily used to assess the growth of the curvature perturbation during ultra-slow-roll inflation for estimating the abundances and the mass ranges of primordial black holes formed when the curvature perturbation re-enters the horizon.

\end{document}